\documentclass[twocolumn,apl,amsmath,amssymb,showpacs,superscriptaddress]{revtex4-2}
\usepackage{epsf}
\usepackage[utf8]{inputenc}
\usepackage{amsmath}
\usepackage{amsfonts}
\usepackage{amssymb}
\usepackage{makeidx}
\usepackage{color}
\usepackage{gensymb}
\usepackage{graphicx}
\usepackage{lipsum}
\usepackage{hyperref}
\usepackage{mathtools,amssymb,lipsum, nccmath}

\begin{document}
\title{Structural and Electrical Transport Properties of NASICON type Na$_{3}$Zr$_{2-x}$Ti$_{x}$Si$_2$PO$_{\rm 12}$ ($x=$ 0.1--0.4) Solid Electrolyte Materials}

\author{Ramcharan Meena}
\affiliation{Department of Physics, Indian Institute of Technology Delhi, Hauz Khas, New Delhi-110016, India}
\affiliation{Material Science Division, Inter-University Accelerator Center, Aruna Asaf Ali Road, New Delhi-110067, India}
\author{Rajendra S. Dhaka}
\email{rsdhaka@physics.iitd.ac.in}
\affiliation{Department of Physics, Indian Institute of Technology Delhi, Hauz Khas, New Delhi-110016, India}

\date{\today}      

\begin{abstract}

We report the structural, resistivity, impedance, and dielectric studies of isovalent substituted Na$_{3}$Zr$_{2-x}$Ti$_{x}$Si$_2$PO$_{\rm 12}$ ($x=$ 0.1--0.4) NASICON type solid electrolyte materials. The Rietveld refinement of XRD patterns shows the monoclinic phase with space group of C 2/c for all the samples. The resistivity analysis shows the Arrhenius-type thermal conduction with an increase in activation energy with doping is explained based on decreased unit cell volume. We use Maxwell-Wagner-Sillars (MWS) relaxation and space charge or interfacial polarization models to explain the frequency and temperature-dependent variations of electric permittivity. The double relaxation peaks in the dielectric loss data show the two types of relaxation mechanisms of different activation energy. The real ($\epsilon^{'}$) and imaginary ($\epsilon^{''}$) parts of permittivity are fitted using the modified Cole-Cole equation, including the conductivity term, which show the non-Debye type relaxation over the measured frequency and temperature range. The impedance analysis shows the contributions from grain and grain boundary relaxation. The fitting performed using the impedance and constant-phase element (CPE) confirm the non-Debye type relaxation. Moreover, the electric modulus analysis confirms the ionic nature having thermally activated relaxation and the modulus scaling analysis shows a similar type of relaxation in the measured temperature range. The modified power law is used to understand the frequency dependence of {\it a.c.} conductivity data. The temperature dependence of exponent ($s$) in modified power law suggests the change in the conduction mechanism from near small polaron tunneling (NSPT) to correlated barrier hopping (CBH) above room temperature. The larger values of  $\epsilon$$_{r}$ indicate these materials as a potential candidate for charge-storage devices.          
 
\end{abstract}

\maketitle
\section{\noindent ~Introduction}

In recent years, sodium (Na) based energy storage devices have attracted great attention due to the low cost as compared to Li-based devices for all solid-state batteries \cite{Jian_AM_17, Singh_Ionics_22}. Among a wide available range, NASICON materials called as Sodium ({\bf Na}) {\bf  S}uper {\bf I}onic {\bf CON}ductor are mostly used as solid electrolyte as well as cathode \cite{SapraAMI24, PatiJPS24} due to their larger ionic conductivity 10$^{-4}$ S/cm at room temperature and 10$^{-1}$ S/cm at 300$^{\circ}$C \cite{Xie_JPCS_16} as well as structural stability \cite{ManishPRB24}. However, the ionic conductivity of these solid electrolytes is still two orders less than the liquid electrolytes, which challenge their commercial usage \cite{Thirupathi_PMS_23, Du_ESM_20}. On the other hand, the liquid-based electrolytes have various disadvantages, such as flammability, leakage, and metallic reduction limits the liquid-based electrolytes for large-area applications. Interestingly, the solid electrolytes have various advantages like high chemical, mechanical, structural, and thermal stability, wide compatibility with materials, large ionic conductivity and high stability in air, making the NASICON-based materials a better choice to use as electrolytes in all solid-state batteries. These materials are also used for various other applications such as solid oxide fuel cells, gas sensors, ion-selective electrodes, supercapacitors, and microwave absorptions, where lower density of NASICON makes them a suitable choice for low-weight devices \cite{Rao_SSI_21, Li_IM_22}. In this line, the NASICON having the chemical composition of Na$_{1+x}$Zr$_2$Si$_x$P$_{3-x}$O$_{\rm 12}$,  (0 $\leq$ $x$ $\leq$ 3) was first characterized by Hong and Goodenough in 1976 \cite{Goodenough_MRB_1976, Hong_MRB_1976}. The original NASICON composition is derived from the NaZr$_{2}$P$_{3}$O$_{12}$ by partially replacing P by Si and compensating the resulting charge difference by adding extra Na inside the matrix. During this process, the lattice parameters along the $c-$axis increase with Si amount for maximum at $x=$ 2 and decrease thereafter.

 Interestingly, the Na$_{1+x}$Zr$_2$Si$_x$P$_{3-x}$O$_{\rm 12}$,  (0 $\le$ $x$ $\leq$ 3) stabilizes in rhombohedral phase (space group R$\bar{3}$C) except 1.8 $\leq$ $x$ $\leq$ 2.2 range which stabilizes in the monoclinic phase (space group C 2/c) \cite{Hong_MRB_1976,  Ran_ESM_21, Goodenough_MRB_1976}. These NASICON structures (monoclinic and rhombohedral) have skeleton-type arrangement containing two ZrO$_{6}$ octahedra separated by three Si/PO$_{4}$ tetrahedra having two different sites available for hopping named as (Na$_{1}$ and Na$_{2}$). The Na$_{1}$ is located between two octahedral sites along the $c-$axis in six-fold coordination with corner-sharing oxygen ions. The Na$_{2}$ is located between two Na$_{1}$ sites along the $a-$axis in eight-fold coordination with O-ions containing four Na sites per unit cell. Here, the rhombohedral phase contains one Na$_{1}$ and three Na$_{2}$ sites having only conduction path (Na$_{1}$-Na$_{2}$-Na$_{1}$). However, the monoclinic phase has one Na$_{1}$, one Na$_{2}$ and two Na$_{3}$ sites as in monoclinic phase Na$_{2}$ site further splits into Na$_{2}$ and Na$_{3}$, which gives the additional conduction path (Na$_{1}$-Na$_{2}$-Na$_{1}$ and Na$_{1}$-Na$_{3}$-Na$_{1}$). Out of these total available sites, 2/3 are filled and 1/3 are available for conduction through the hopping of Na ions between the sites \cite{Naqash_SSI_18, Park_JPS_18}.  The Na ions migrate  through an oxygen triangle (made from Na and O ions of two polyhedrons) called as bottleneck. The conductivity depends on the number of ions available for conduction and the  bottleneck area. Here, the highest conductivity at room temperature is found for the Na$_{3}$Zr$_{2}$Si$_2$PO$_{\rm 12}$ ($x=$ 2) sample which correlate with its larger $c-$axis value \cite{Rao_SSI_21, Naqash_SSI_18, Park_JPS_18}. Notably, the monoclinic phase transforms to the rhombohedral phase at around 160$^{\circ}$C has been confirmed using in-situ x-ray diffraction (XRD) and differential scanning calorimetry measurements (DSC) \cite{Park_JPS_18, Jolley_JACS_15}. In this case, only Na ion ordering changes during the phase transition where one Na$_{2}$ and two Na$_{3}$ sites of the monoclinic phase merge into three Na$_{2}$ sites of the rhombohedral phase \cite{Park_JPS_18}. The analysis shows that the transformation from monoclinic to rhombohedral phase is due to the shear deformation of the unit cell \cite{Park_JPS_18, Jolley_JACS_15}. 

Furthermore, the conductivity of NASICON based solid electrolyte materials can be tuned by changing the synthesis method, amount of initial precursors, sintering temperature, doping, valance state of dopant, Si to P ratio and the radius of dopant ions \cite{Wang_NS_22, Ruan_CI_19, Jiang_CEJ_23, Wang_ESM_23, Oh_AMI_19, Lu_AEM_19, Sun_AFM_21, Khakpour_ECA_16}. It is concluded from previous  studies that lower sintering temperatures during sample preparation and a larger density of prepared NASICON samples are preferred for various applications like solid electrolyte. The doping of divalent or trivalent atoms at the Zr site will change the number of free charge carriers per unit cell. The dopant's ionic radius (if larger than Zr ion) increases the bottleneck area, which decreases the activation energy resulting an overall increase in Na conduction. There are very few dielectric studies on doped NASICON materials, and limited to above room temperature and in the microwave region only \cite{Dubey_AEM_21, Chen_ML_18, Chen_JECS_18}. In recent, we have studied the pristine \cite{Meena_CI_22}, divalent (Ni$^{2+}$) \cite{Meena_PB_24}, trivalent (Pr$^{3+}$) \cite{Meena_JACS_24} doped NASIOCN samples in the temperature range of 90--400~K and  frequency range of 20 Hz -- 2 MHz. However, no reports are available to the best of our knowledge on the isovalent doping to understand the dielectric properties and their relaxation mechanism. 

Therefore, in this paper, we have prepared the isovalent doped Na$_{3}$Zr$_{2-x}$Ti$_{x}$Si$_2$PO$_{\rm 12}$ ($x=$ 0.1--0.4) NASICON materials using solid-state reaction method. We have chosen Ti ion (53 pm) for isovalent doping at the Zr ion (72 pm) site. The Ti being iso-valent to Zr will not introduce additional Na ions at the interstitial sites and lower the unit cell volume, which helps in increased polarization in the presence of external electric fields. The structural studies are performed to confirm the crystal structure and the {\it d.c.} resistivity measurements are done to understand the conduction mechanism. The dielectric, impedance, and a.c. conductivity analysis are performed at various temperatures and wide frequency ranges. The experimentally obtained data are fitted using various models to find the nature of relaxation and conduction mechanisms.

\section{\noindent ~Experimental}

The Na$_{3}$Zr$_{2-x}$Ti$_{x}$Si$_2$PO$_{\rm 12}$ ($x=$ 0.1--0.4) samples are prepared using the solid-state reaction method by taking Na$_{3}$PO$_{4}$.12 H$_{2}$O (purity 99.5 \%), ZrO$_2$ (purity 99.5 \%), TiO$_2$ (purity 99.8 \%) and SiO$_2$ (purity 99\%) in stoichiometric amount. The SiO$_2$ powder was preheated at 200$^\circ$C for 12 hrs to remove the moisture. The Na$_{3}$PO$_{4}$.12 H$_{2}$O is taken in 15\% excessive amounts to compensate for the loss of Na and P ions during the high-temperature sintering process. All the precursors were thoroughly grounded using the Agate mortar pestle for uniform mixing. The mixed powder after calcination at 1100$^\circ$C for 12 hrs became sky blue in color and obtained in colloidal form. This powder was again ground to have a uniform distribution of particles and pressed into pellets of diameter 10~mm with a thickness of 1~mm using the hydraulic pressure (1100 PSI) with a holding time of 5 mins. These prepared pellets (sky blue color) were sintered at 1150$^\circ$C for 8 hrs. The heating and cooling rates during the calcination and sintering were fixed at $5^\circ$C/min.   

   \begin{figure*}
    \centering
    \includegraphics[width=0.83\textwidth]{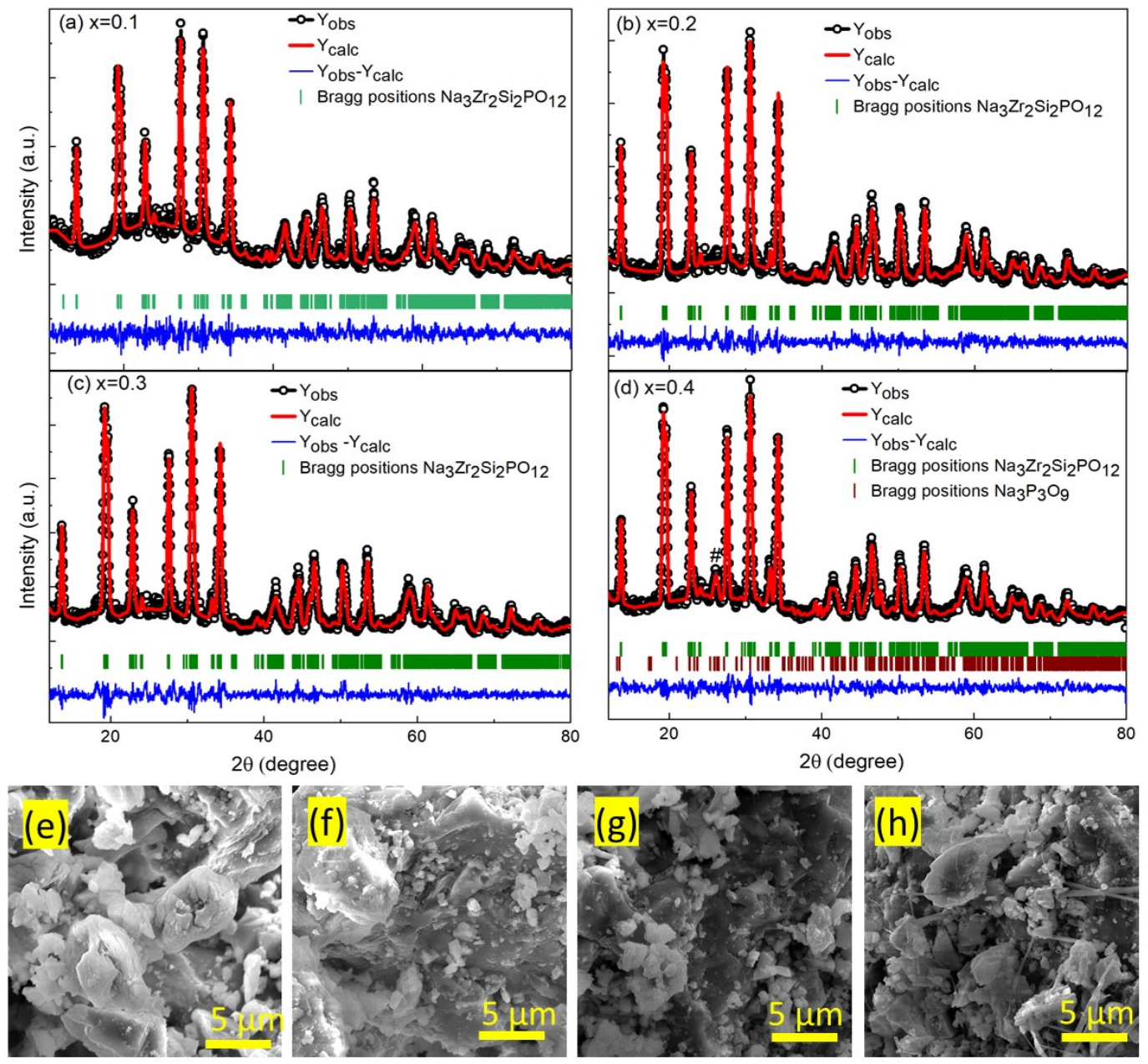}
    \caption{(a--d) The Reitveld refined XRD patterns of Na$_{3}$Zr$_{2-x}$Ti$_{x}$Si$_2$PO$_{\rm 12}$ ($x=$ 0.1--0.4) samples, where open black and red solid lines represent the experimental data and refined profile, green vertical lines show Bragg positions of the monoclinic NASICON phase, and Bragg's position corresponds to Na$_{3}$P$_{3}$O$_{9}$ impurity phase are marked by \# symbols are shown by wine color in (d). The blue solid line represents the difference between observed and calculated data. (e--h) The FE-SEM images obtained by taking the acceleration voltage of 15 kV with a scale size of 5 $\mu$m. }
\label{XRDSEM}
 \end{figure*}

The X-ray diffraction (XRD) measurements are performed in the 2$\theta$ range of 10$^\circ$-80$^\circ$ with a step size of 0.0167$^\circ$ using PANalytical X’Pert-PRO diffractometer having Cu-K$_\alpha$ radiation of wavelength ($\lambda$=1.54~$A^\circ$) in Bragg-Brentano geometry. The FullProf suite software is used for the Rietveld refinement of recorded XRD patterns of all the samples to determine the crystal structure and phase symmetry \cite{Carvajal_CRNS_00}. The microscopic analysis was performed using the MIRA II LMH field emission scanning electron microscope (FESEM). The samples were fixed on the circular holder using double-sided carbon tape. Since these samples are insulating type in nature, a coating of Au-Pd alloy (thickness 5~nm) is performed before starting the experiment. The elemental analysis is performed using NCA PentaFET3 energy dispersive X-ray detector from Oxford attached with SEM. The temperature-dependent electrical resistivity measurements are performed using a 6517 B Keithley electrometer by applying 1~V excitation voltage and measuring the resultant current. Various frequency-dependent parameters like {\it(Capacitance, dielectric loss, impedance, and phase angle)} are measured in parallel capacitance mode by the standard four-probe method using a LCR meter (Model--E4980A) from Agilent. These measurements are performed in the frequency range of 20 Hz--2~MHz keeping 1~V {\it a.c.} signal as an input perturbation. Open, short, load and cable length corrections are performed before starting the frequency-dependent measurements to avoid any error arise due to cable capacitance and conductance. The metallic electrodes on the NASICON samples are fabricated by coating the pellets using silver paint and dried at 200$^\circ$C for 2 hrs. The edge effect of the electrodes is minimized by keeping the electrode edge inside the sample boundary. The temperature-dependent measurements are performed using the Lakeshore controller (Model-340) in the range of 100--440~K with a fixed heating rate of 2$^\circ$C/min having  stability of 100~mK and a time stability of 2 mins. All temperature-dependent measurements are performed by dipping the sample into liquid nitrogen Dewar for uniform cooling, maintaining a vacuum around $10^{-3}$ mbar.

\section{\noindent ~Results and discussion}

The room temperature Rietveld refined XRD patterns of Na$_{3}$Zr$_{2-x}$Ti$_{x}$Si$_2$PO$_{\rm 12}$ ($x=$ 0.1--0.4) samples are shown in Figure~\ref{XRDSEM}(a--d). The refinement has been processed using FullProf suite software \cite{Carvajal_CRNS_00} by taking the pseudo-voigt peak shape and linear interpolation is used for background correction. The scale factor, FWHM parameters ($u$, $v$, $w$), lattice parameters ($a$, $b$, $c$, $\beta$), shape parameters (Eta, $x$), atomic positions ($x$, $y$, $z$), and asymmetry are considered as free parameters; whereas, the occupancy factor was kept fixed during the refinement process. The excellent matching between the calculated and experimental data having reduced $\chi$$^2$ values in the range of 1.43 to 1.50 confirms the monoclinic phase with C 2/c space group for all the samples with a small amount of Na$_{3}$P$_{3}$O$_{9}$ impurity for $x=$ 0.4 sample, as denoted by \# symbol in Figure~\ref{XRDSEM}(d). The obtained lattice parameters are presented in Table 1 of ref.~\cite{SI_RCM_24}, which agree with the values reported in refs.~\cite{Wang_JPCS_23, Jolley_JACS_15}. We observe that all parameters ($a$, $c$, $\beta$ and volume) decrease, while $b$ nearly remains constant with Ti doping. This type of variation in lattice parameters is expected as  the ionic radius of the dopant element Ti (53 pm) is lower than the Zr (72 pm) ion, that results in a decrease of lattice parameters and unit cell volume with an increase in doping concentration. Moreover, the FE-SEM images shown in Figures~\ref{XRDSEM}(e--h) confirm the agglomerated type growth for the samples with lower doping and the dense microstructure for the samples with higher doping. This indicates that the density of the microstructure increases with doping, with some nano rod-like structures for the highest doped samples. This rod-like structure is due to Na$_{3}$P$_{3}$O$_{9}$ impurity phase, as confirmed by the energy-dispersive x-ray (EDX) analysis. The elemental analysis using EDX analysis also confirms the presence of constituent elements (Na, Zr, Ti, Si, P, and O) in desired stoichiometry for all the samples. 
       
\begin{figure}[h]
    \centering
    \includegraphics[width=0.47\textwidth]{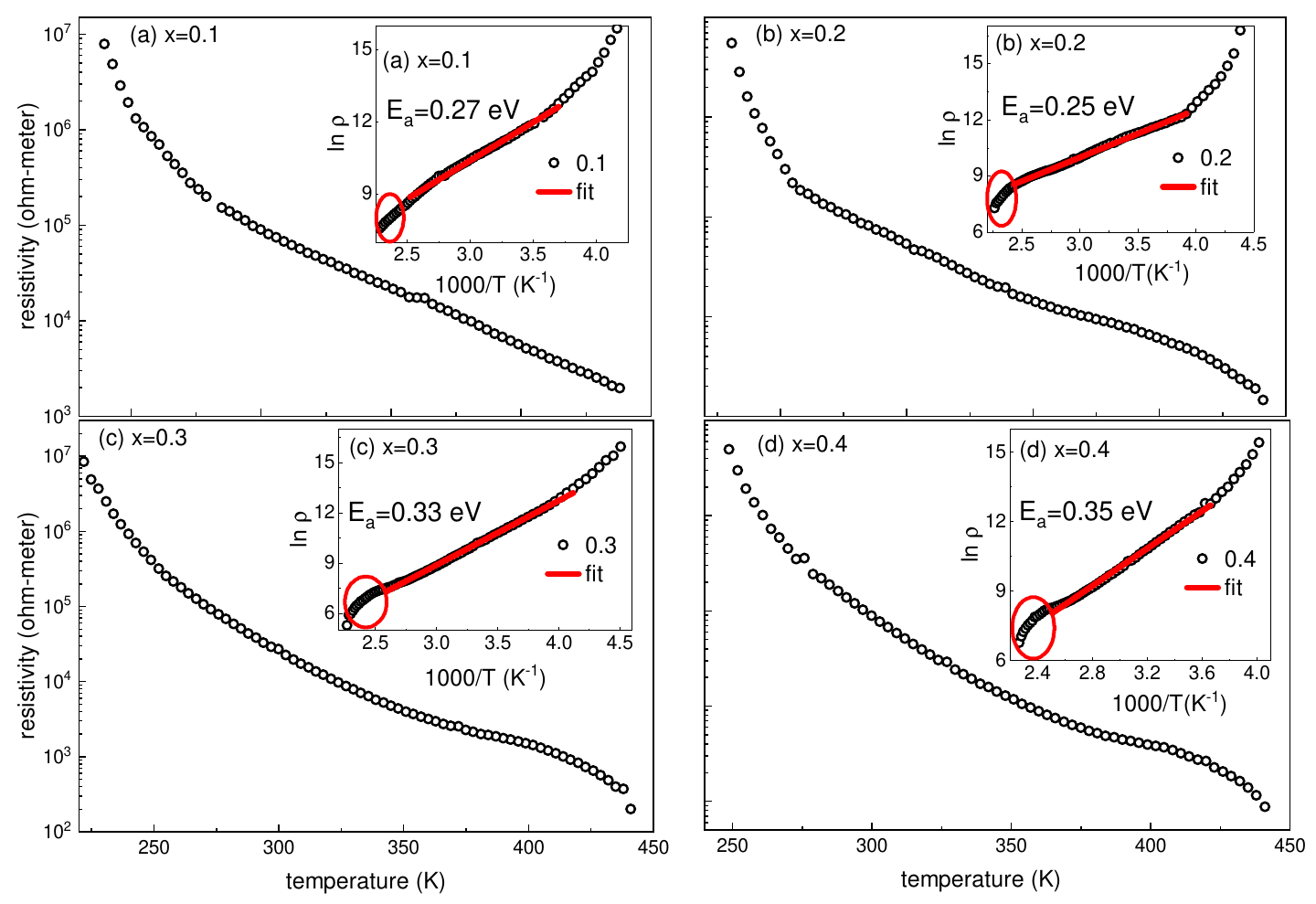}
    \caption{The temperature dependent resistivity variations of the Na$_{3}$Zr$_{2-x}$Ti$_{x}$Si$_2$PO$_{\rm 12}$ ($x=$ 0.1--0.4) samples. The inset shows the ln $\rho_{\rm T}$ vs ($\frac{1000}{T}$) graph where the open symbol represents the experimental data, and the solid line shows the linear fit using the Arrhenius thermal conduction model.} 
\label{RT}
 \end{figure}
 
The temperature-dependent resistivity variation, in Figures~\ref{RT}(a-d), shows a semiconducting behavior within the measured temperature range with a strong insulating nature below 250 K. The resistivity decreases with increasing temperature due to the enhanced mobility of charge carriers and thermal activation of charge carriers. The activation energy of charge carriers is determined using the Arrhenius thermal conduction model as given by \cite{Kumar_JVSTA_23, Meena_CI_22, Meena_PB_24}  
\begin{equation}
    \rho_{T}=\rho_{0} ~exp(\frac{-E_a}{k_B T})
\end{equation} 
Here, $E_a$ is the thermal activation energy, $k_B$ is the Boltzmann constant, T is the measuring temperature, $\rho_0$ is the pre-exponential factor or called resistivity at zero temperature and $\rho_T$ is the resistivity measured at temperature T. The Arrhenius thermal activation energy is determined using the slope of ln~$\rho$$_{T}$ vs ($\frac {1000}{T}$), and the activation energy values are found in the range of 0.25 to 0.35 eV (increases with doping) for all doped samples. The activation energy increases with doping as the lower ionic radius of Ti ions increases the interaction among the charge carriers. The interaction becomes stronger with doping due to a decrease in unit cell volume (confirmed by XRD), requiring more thermal energy for conduction, resulting in an increased magnitude of activation energy with doping. The experimental data could not be fitted above 400~K indicating the deviation, as shown by the red circle in the inset of Figures~\ref{RT}(a--d) from the Arrhenius model at higher temperatures. The deviation indicates a signature of structural phase transition from monoclinic to rhombohedral phase around 420~K.    

\begin{figure*}
    \centering
    \includegraphics[width=0.85\textwidth]{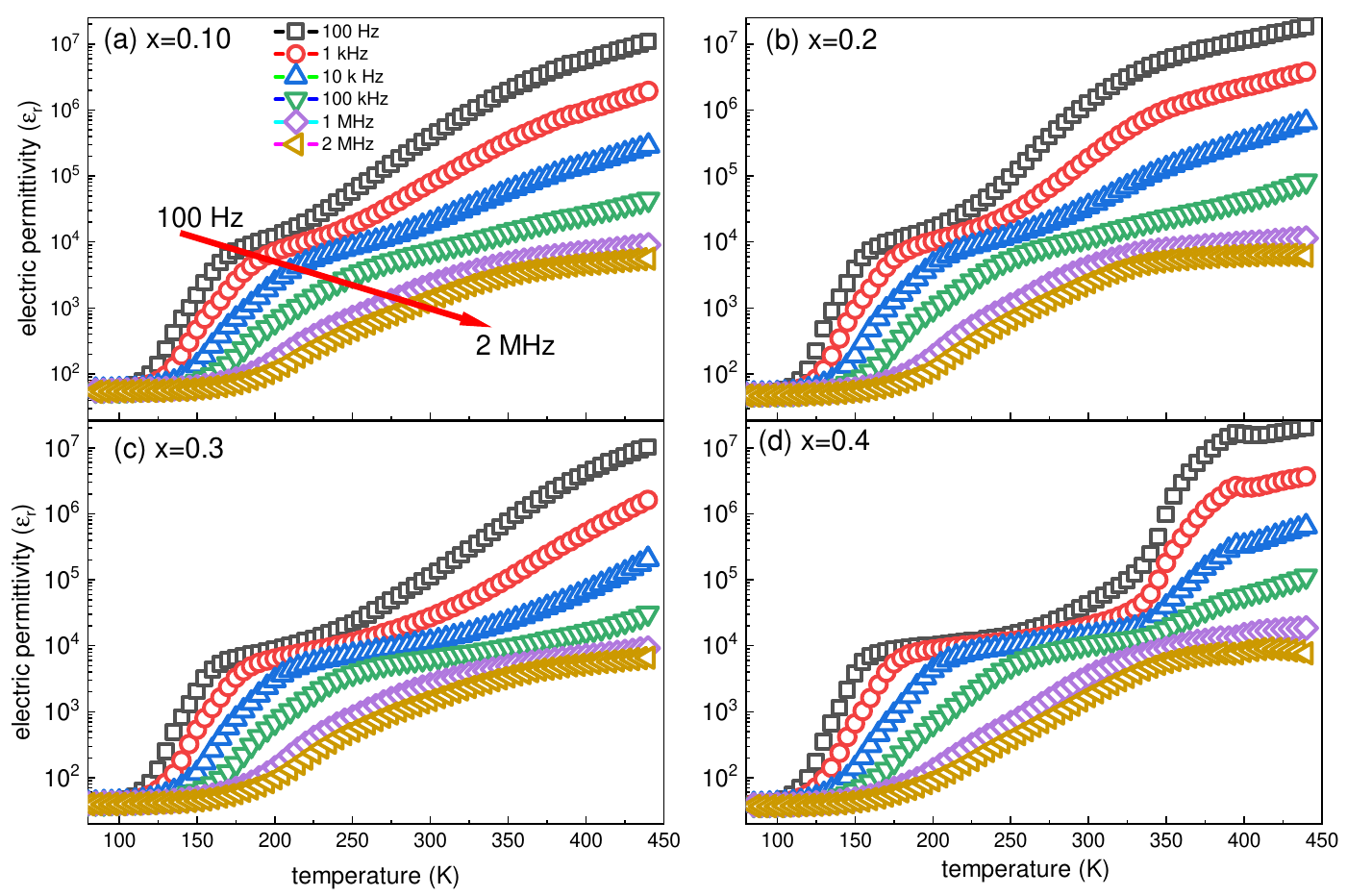}
\caption {The dielectric permittivity with temperature at selected frequencies for Na$_{3}$Zr$_{2-x}$Ti$_{x}$Si$_2$PO$_{\rm 12}$ ($x=$ 0.1--0.4) samples. The arrow in (a) shows the direction of increasing frequency and shifting the relaxation peak towards the high-temperature.}
 \label{CT}
\end{figure*}

Figures~\ref{CT}(a--d) shows the dielectric permittivity plotted with temperature at selected frequencies, which shows a smaller magnitude (of the order 10$^{1}$) at lower temperatures and reaches the magnitude (10$^{7}$) around 400 K for 100 Hz, suggesting the possible application of these samples as high K dielectric materials. The step-like increase of dielectric permittivity with temperature indicates the relaxation-type behavior for all the samples, shifts towards the high-temperature side with an increase in frequency, as shown by an arrow in Figure~\ref{CT}(a). This type of relaxation behavior can be explained using the Maxwell-Wagner-Sillars (MWS) relaxation and space charge or interfacial polarization models. According to the MWS model, a polycrystalline material is considered to be made of well-conducting grains separated by poorly conducting grain boundaries. Under the influence of an externally applied electric field, the charge carriers are trapped at the grain boundaries, giving enhanced polarization. Similarly, the charge carriers are trapped at the interface of the sample and metal electrode, called interfacial or space charge polarization. The temperature and frequency dependence of dielectric permittivity can be explained as follows: (a) If we consider the effect of temperature at a fixed frequency, at lower temperatures, the charge carriers do not have sufficient thermal energy to orient in the direction of the applied field gives the smaller magnitude of dielectric permittivity. As we increase the temperature, the average thermal energy available with carriers increases, enhancing carriers' mobility and generating the additional carriers, producing a higher amount of polarization and resulting in larger dielectric permittivity. (b) Let's consider the effect of frequency at a fixed temperature for lower frequencies where the charge carriers can follow the externally applied electric field and create a potential barrier inside the sample, which leads to the piling of the charge carriers at the grain boundary (MWS polarization) or sample-electrode interface (interfacial polarization), producing a higher amount of polarization, resulting in a larger dielectric permittivity. At higher frequencies, the field variations are so rapid that carriers are no longer able to follow the externally applied field (due to the shorter time period available with carriers), giving a decrease in polarization produces smaller values of dielectric permittivity. 

\begin{figure*}
    \centering
    \includegraphics[width=0.8\textwidth]{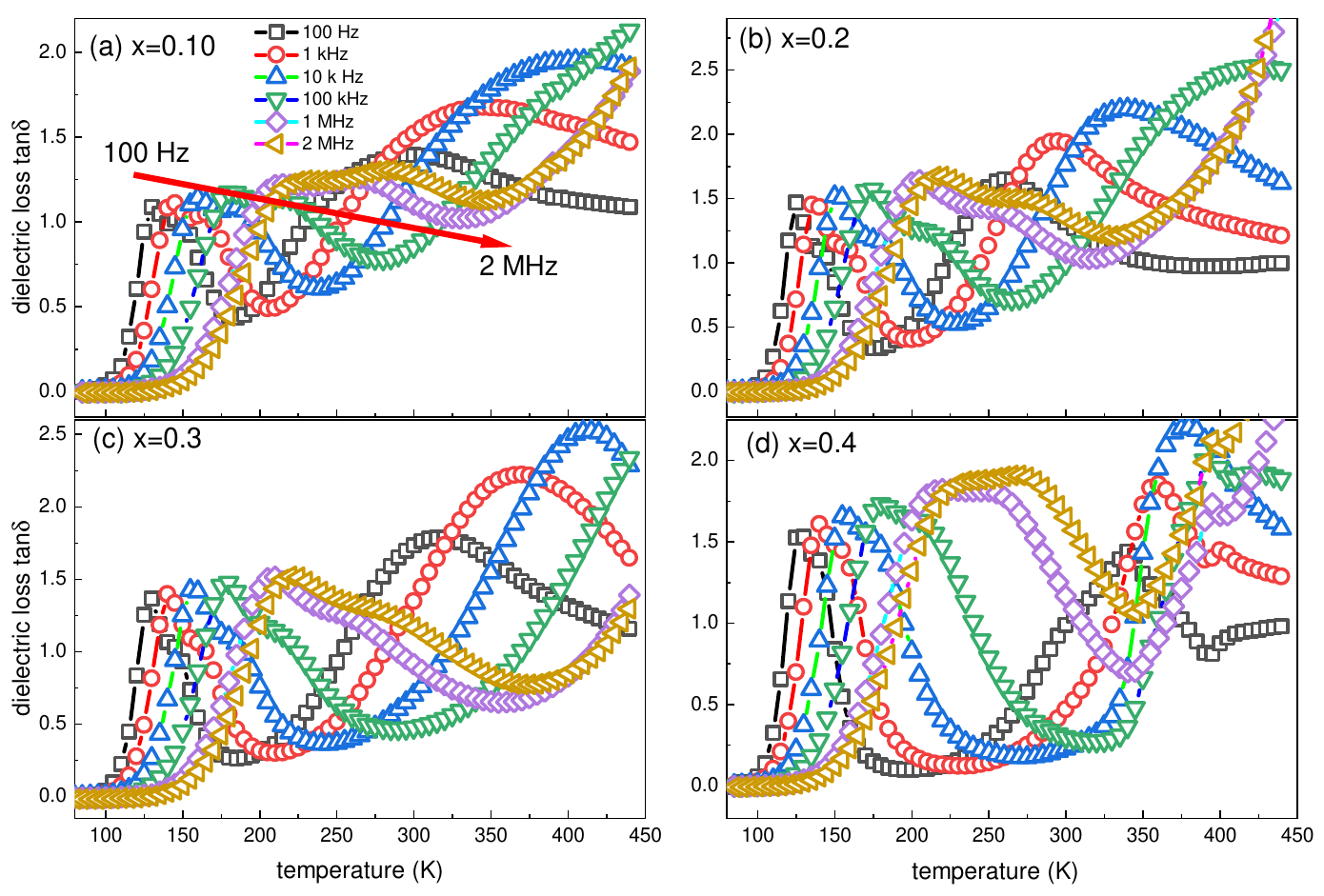}
    \caption{The dielectric loss variations for Na$_{3}$Zr$_{2-x}$Ti$_{x}$Si$_2$PO$_{\rm 12}$ ($x=$ 0.1--0.4) samples at selected frequencies as a function of temperature. The arrow shows the direction of increasing frequency along with the shift in the position of relaxation peak towards the high-temperature.}
\label{DT}
 \end{figure*}

Furthermore, we present the dielectric loss variation at selected frequencies in Figures~\ref{DT}(a--d), which show a rapid decrease in dielectric permittivity resulting a peak in dielectric loss spectra called a relaxation peak. The dielectric loss appears due to a lag in polarization compared to the applied field where the total loss inside dielectric materials is the combined effect of relaxation loss and conduction loss. The relaxation loss is generally dominated at the higher frequency side where the carriers no longer follow the applied field due to the fast change in the applied field. The rapid friction, that occurs at high frequencies and appears as heat inside the dielectric materials, is called as relaxation loss. We find that the height of the relaxation peak is increased with an increase in frequency for all the samples, indicating the relaxation loss is dominated at higher frequencies. The conduction loss mainly occurs due to the finite conductivity of the sample being dominated at lower frequencies where free carriers such as vacancies, interstitials, and point defects follow the external field, producing heat inside the materials due to scattering among the carriers. The increase in temperature enhances the scattering among the carriers as well as generates new free carriers, resulting in increased conduction in the dielectric materials, hence the dielectric loss. 
The peak in the dielectric loss appears once the applied field frequency matches the carrier's hopping frequency, where the externally applied energy is transferred to the dielectric materials. We observe that the peak width increases with frequency due to the spread in relaxation time (related by $\omega$$\tau$=1). Also, the peaks shift towards the high-temperature side with frequency, indicating the thermal activation of the relaxation peak. The relaxation time and its temperature dependence are given by \cite{Ang_PRB_00, Jia_JAP_11, Meena_CI_22}
\begin{equation}
     \tau = \tau_0 ~ exp~(\frac{E_{a}}{k_B T_m})
\label{tau}
 \end{equation}
Here, the $\tau$ is measured relaxation time, $\tau_0$ is the characteristic relaxation time at infinity temperature, $E_{a}$ is the activation energy of dielectric relaxation,  k$_B$ is Boltzman constant and $T_m$ is the measured peak temperature. The activation energy ($E_{a}$) and characteristic relaxation time ($\tau_0$) are determined using the ln($\tau$) vs ($\frac{1000}{T}$) graphs, where the slope of the graph provides the activation energy ($E_{a}$) and intercept will provides the characteristic relaxation time ($\tau_0$). Interestingly, the dielectric loss spectra in Figures~\ref{DT}(a--d) show the double relaxation peaks, which indicate two types of activation energy in the measured temperature range having a magnitude of 0.24--0.26 eV and 0.40--0.51 eV for the low and high-temperature relaxation peaks, respectively, as shown in Figure 2 of \cite{SI_RCM_24}. The magnitude of activation energy suggests the polaron-type hopping conduction in the measured temperature range \cite{Rehman_JALCOM_14, Tailor_JPCC_22}. In this hopping type, it creates structural disorder in its surrounding medium by moving the atoms from its equilibrium state. The obtained characteristic relaxation time ( $\tau_0$) is in the range of 1.17--10.39$\times$10$^{-13}$ sec for Peak 1 and 1.13--1.86$\times$10$^{-11}$ sec for Peak 2.          

\begin{figure*}
    \centering
    \includegraphics[width=0.8\textwidth]{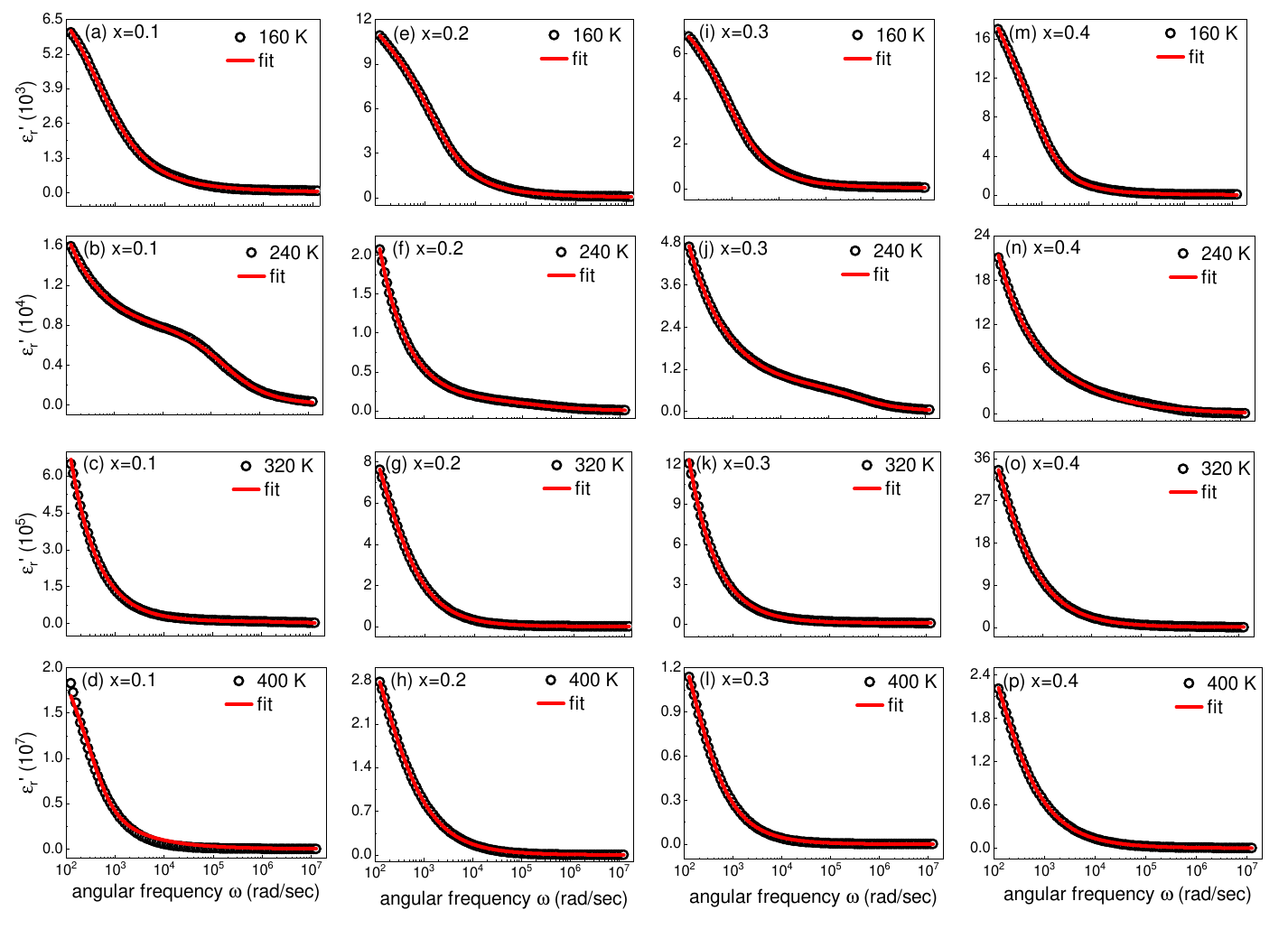}
    \caption{The real part of electric permittivity ($\epsilon$ $^{'}$) variations with frequency at selected temperatures for the Na$_{3}$Zr$_{2-x}$Ti$_{x}$Si$_2$PO$_{\rm 12}$ ($x=$ 0.1--0.4) samples. The open symbols are observed in experimental data, and a solid line is fit using the modified Cole-Cole equation by including the conductivity term \ref{E'}.}
    \label{RE}
\end{figure*}

\begin{figure*}
    \centering
    \includegraphics[width=0.77\textwidth]{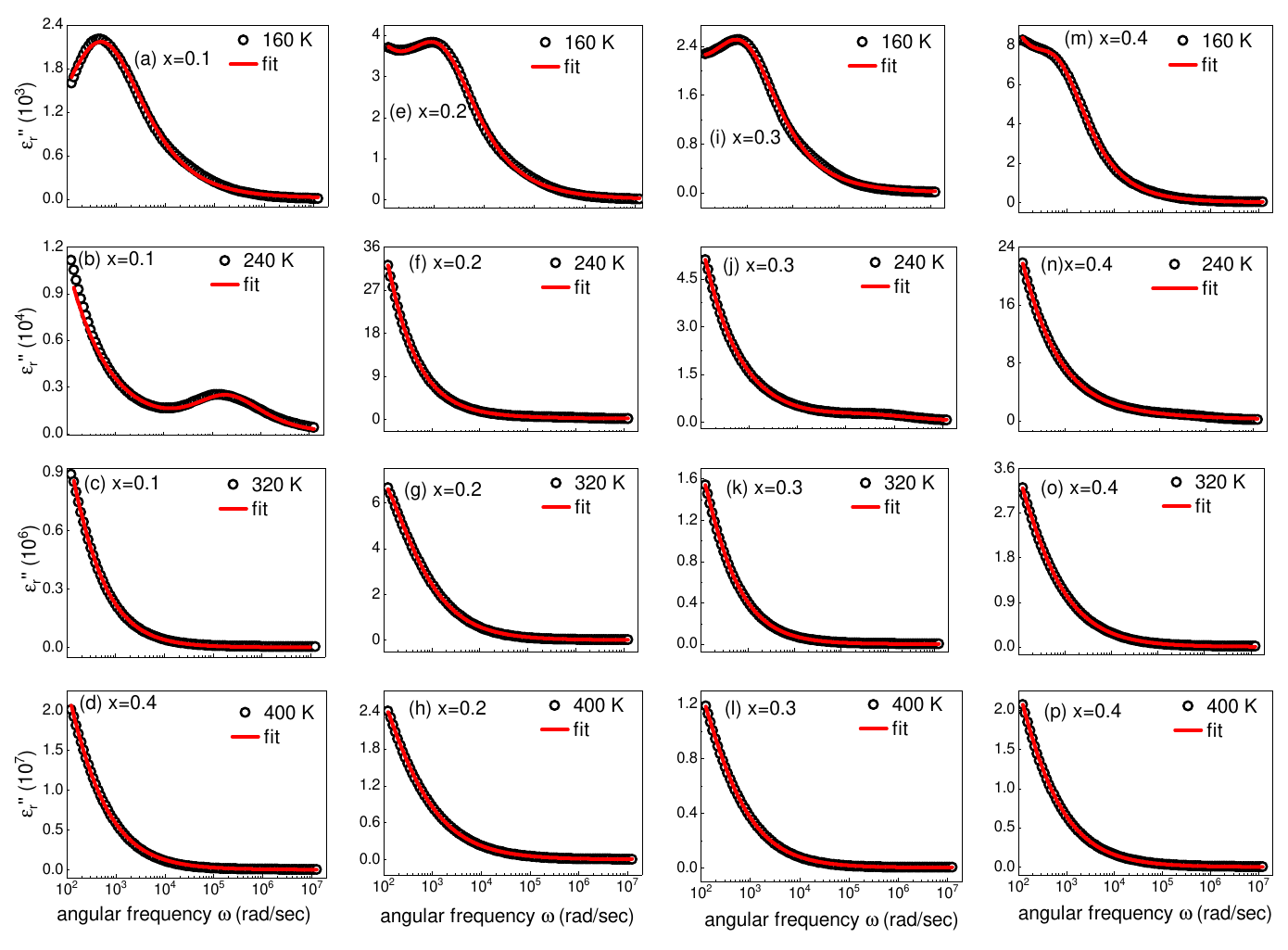}
    \caption{The imaginary part of electric permittivity ($\epsilon$ $^{''}$) variations with frequency at selected temperatures for the Na$_{3}$Zr$_{2-x}$Ti$_{x}$Si$_2$PO$_{\rm 12}$ ($x=$ 0.1--0.4) samples. The open symbols are observed in experimental data, and a solid line is fit using the modified Cole-Cole equation by including the conductivity term \ref{E''}.}
    \label{IE}
\end{figure*}

To further understand the frequency-dependence of permittivity and relaxation behavior in detail, isothermal curves of the real ($\epsilon^{'}$) and imaginary ($\epsilon^{''}$) parts of dielectric permittivity are shown in Figures~\ref{RE} and \ref{IE}, respectively. A similar type of variation is observed for all Ti doped NASICON samples where the dielectric permittivity increases with temperature and follows the inverse behavior with frequency. The larger increase in permittivity at lower frequencies is due to the increased polarization. In the low-frequency region, the charge carriers follow the applied field direction and pile up at the sample electrode interface (space charge polarization) or boundary inside the bulk (Maxwell-Wagner polarization) by creating a high energy barrier in the field direction. This piling of charges increase the polarization inside the materials, hence the dielectric permittivity. At higher frequencies, the field variations are so rapid that the charge carriers not able to follow the applied external field, decreasing the polarization, which results in a decrease in the real part of dielectric permittivity ($\epsilon^{'}$) \cite{Nallamuthu_JALCOM_11}. The permittivity increases with temperature due to an increase in the number of carriers piling at the interface, giving the larger polarization inside the materials, resulting in an increase in permittivity. 

Moreover, in Figure~\ref{IE}, the relaxation peaks are observed in imaginary permittivity data ($\epsilon^{''}$) shifts towards the high-frequency side due to higher rate of polarization with an increase in temperature. The broad nature of the relaxation peaks suggesting a distribution of relaxation times in the measured frequency and temperature range. The dispersive nature of dielectric relaxation can be determined using the Cole-Cole equation \cite{Cole_JCP_1941, Macdonald_Wiley_05} 
\begin{equation}
    \epsilon^*=\epsilon^{'} - \iota \epsilon^{''}=\epsilon_{\infty} + \frac {\epsilon_s-\epsilon_{\infty}}{[1+ (\iota \omega \tau)^{1-\alpha}]}
\label{Cole}
\end{equation}
Here, $\epsilon$$_s$-$\epsilon$$_\infty$ is the dielectric strength of the material, $\tau$ is the mean relaxation time, and $\alpha$ is the Cole-Cole parameter having values between 0 to 1 provides the information about the type of interaction among the charge carriers. For ideal Debye-type situation considers no interaction among the charge carriers having the single relaxation time with $\alpha=$ 0. However, if there is significant interaction among the charge carriers, a broad peak shape is observed, containing the distribution of relaxation times having $\alpha$$>$0. We have tried to fit our data using equation \ref{Cole}; the data is well-fitted in the high-frequency region. However, we are unable to fit the data in the low-frequency region. From the literature, it is found that the low-frequency data can be fitted by adding the {\it d.c.} conduction contribution in the relaxation term. The modified Cole-Cole equation, including the conductivity term, is given by \cite{Chanda_MRB_13, Thongbai_JPCM_08}       
\begin{equation}
    \epsilon^*=\epsilon^{'} - \iota \epsilon^{''}=\epsilon_{\infty} + \frac {\epsilon_s-\epsilon_{\infty}}{[1+ (\iota \omega \tau)^{1-\alpha}]} + \frac{\sigma^{*}}{\iota \omega^s}
\label{epsilon}
\end{equation}
Here, $\sigma$$^{*}$ is the complex conductivity term. Using the above relation, the real ($\epsilon^{'}$) and imaginary part ($\epsilon$ $^{''}$) of the complex permittivity ($\epsilon^{*}$) are given by 
\begin{subequations}
\begin{equation}
\epsilon^{'}= \epsilon_{\infty}+ \frac{(\epsilon_s-\epsilon_\infty)[1+(\omega\tau)^{1-\alpha} sin(\frac{\alpha \pi}{2})]}{[1+2 (\omega\tau)^{1-\alpha}  sin(\frac{\alpha \pi}{2})+ (\omega\tau)^{2-2\alpha}]}+\frac{\sigma_2}{\epsilon_0 \omega^s}
\label{E'}
 \end{equation}  
\begin{equation}
\epsilon^{''}= \frac{(\epsilon_s-\epsilon_\infty)[(\omega\tau)^{1-\alpha} cos(\frac{\alpha \pi}{2})]}{[1+2 (\omega\tau)^{1-\alpha}  sin(\frac{\alpha \pi}{2})+ (\omega\tau)^{2-2\alpha}]}+\frac{\sigma_1}{\epsilon_0 \omega^s}  
\label{E''}
 \end{equation}  
\end{subequations}
In the above equation, $\sigma_1$ is the conductivity contributions due to free charger carriers or d.c. conductivity, and $\sigma_2$ is the contributions from the bound charge carriers or localized carriers, and $s$ is an exponent having the values between 0 and 1. In the case of ideal complex conductivity, $s$~=1. If s~$<$1, the polarization process has the distribution of various relaxation processes. As shown in the above equations, the bound charge carriers increase the contribution in the charge storage ($\epsilon^{'}$), and the free charge carrier's contributions increase the dielectric loss inside the materials ($\epsilon^{''}$). The first term in equation \ref{E''} shows the contributions from permanent dipoles or the carriers that have short-range movement of carriers and the second term includes the contributions from carriers having long-range migration of carriers. The isothermal frequency dependence of real ($\epsilon^{'}$) and imaginary ($\epsilon^{''}$) permittivity are well fitted using the equations~\ref{E'} and \ref{E''}. The fitting results suggested that the dipolar and conductivity relaxations are mainly responsible for the dielectric relaxation in these samples. The value of the $s$ parameter approaches towards one at high temperatures, indicating that polarization is weakly dispersive  \cite{Chanda_MRB_13, Thongbai_JPCM_08}.    

\begin{figure*}
    \centering
    \includegraphics[width=0.8\textwidth]{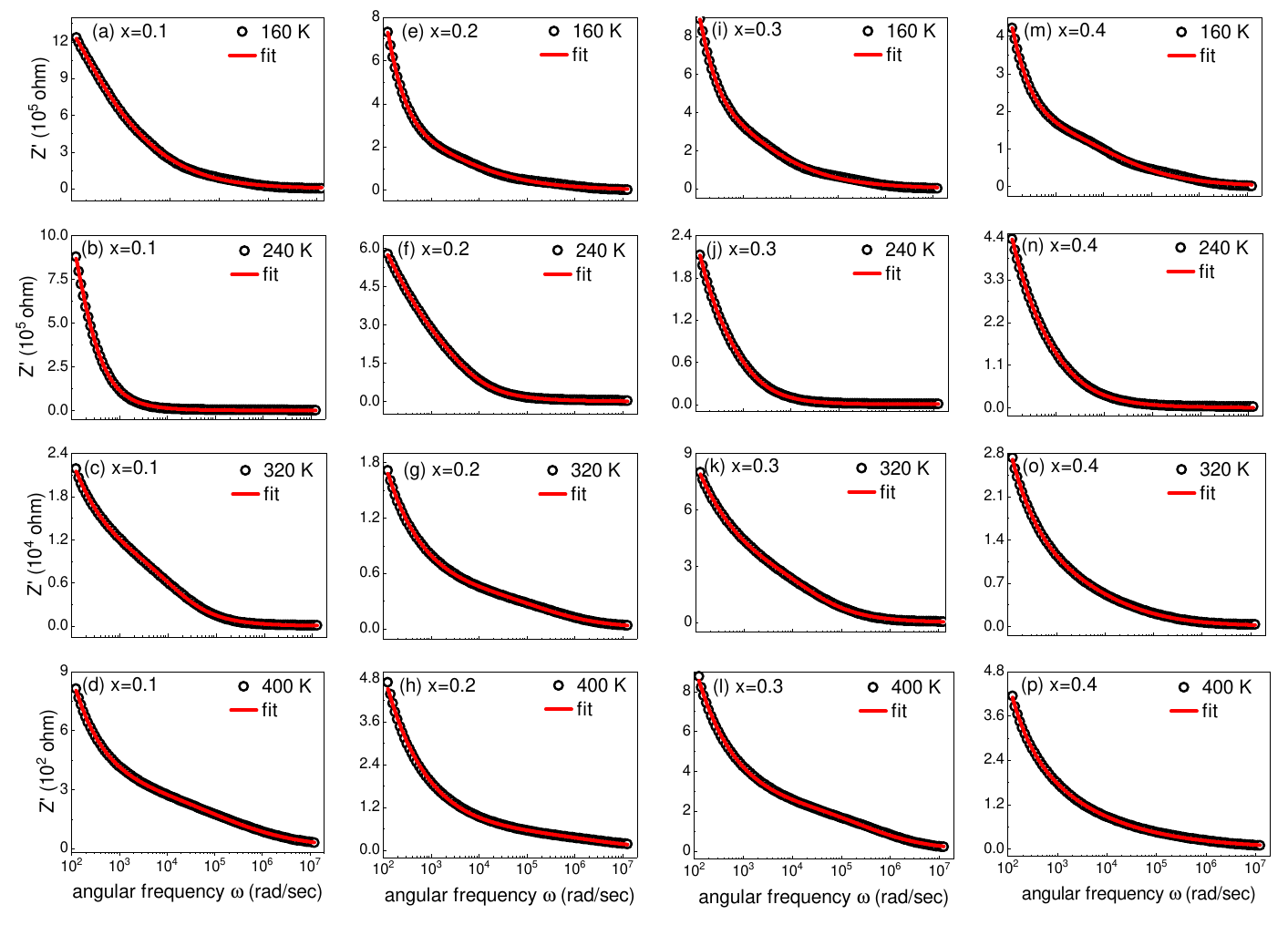}
    \caption{The real part (Z$^{'}$) of total impedance (Z) variations as a function of frequency for the Na$_{3}$Zr$_{2-x}$Ti$_{x}$Si$_2$PO$_{\rm 12}$ ($x=$ 0.1--0.4) samples at selected temperatures. Here, the open symbol represents the observed experimental data, and the solid line is fit using equation \ref{RZ}.}
\label{ReZ}
 \end{figure*}

\begin{figure*}
    \centering
    \includegraphics[width=0.8\textwidth]{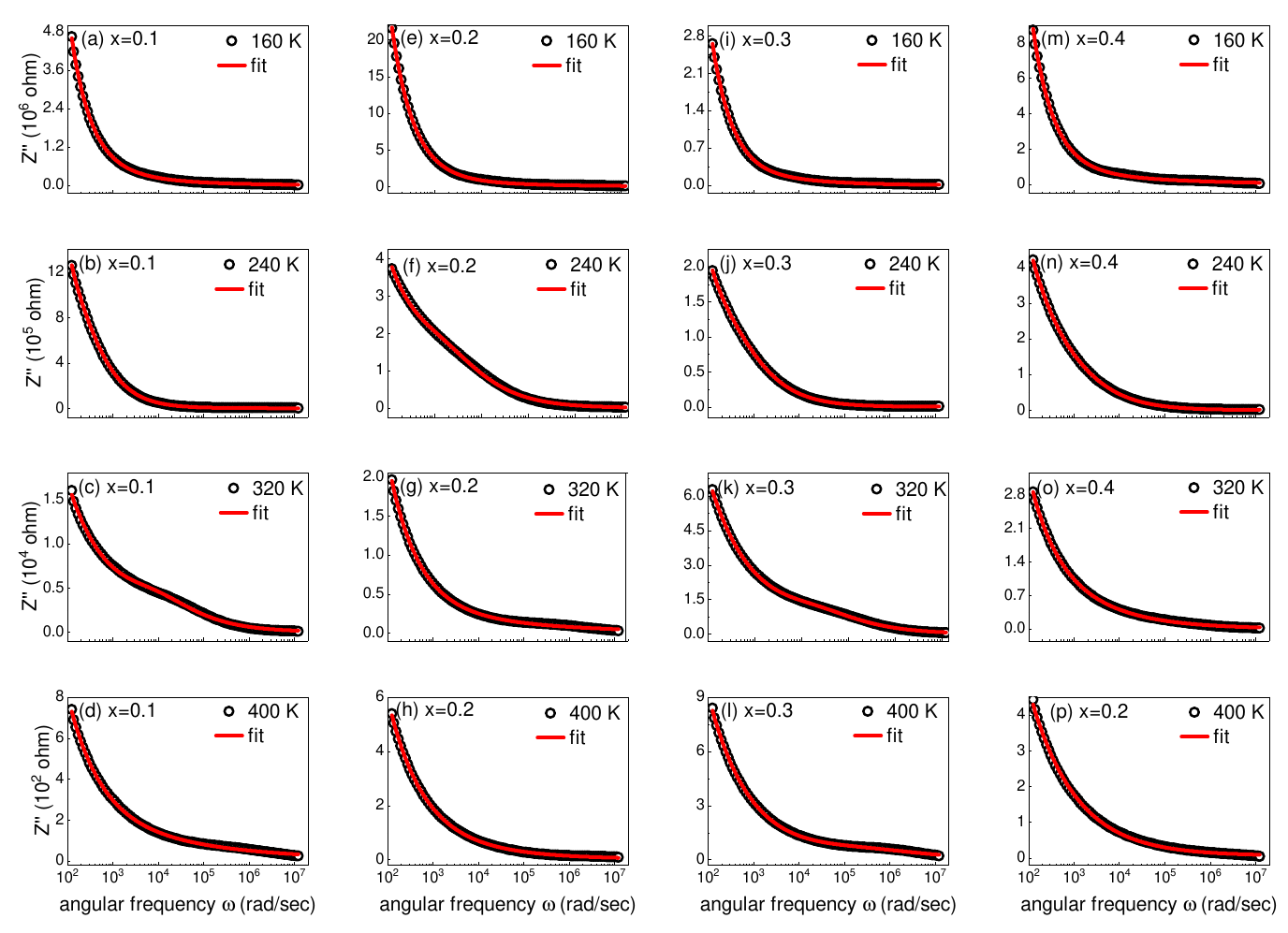}
    \caption{The imaginary part (Z$^{''}$) of total impedance (Z) variations as a function of frequency for the Na$_{3}$Zr$_{2-x}$Ti$_{x}$Si$_2$PO$_{\rm 12}$ ($x=$ 0.1--0.4) samples at selected temperatures. Here, the open symbol represents the observed experimental data, and the solid line is fit using equation \ref{IZ}.}
\label{ImZ}
 \end{figure*}

It is important to note that the impedance spectroscopy is an important tool to determine the type of relaxation, contributions of various relaxation mechanisms as a function of temperature  and frequency using the real (Z$^{'}$) and imaginary (Z$^{''}$) part of impedance, as shown in Figures~\ref{ReZ} and \ref{ImZ}, respectively. The impedance data are normalized using the geometrical factor $g$=$\frac{A}{2d}$, where A is the electrode area and $d$ is the thickness of the pellet \cite{Chandrasekhar_JPCM_12, Schmidt_PRB_09}. The real (Z$^{'}$) and imaginary (Z$^{''}$) parts of impedance are measured by applying an alternating voltage signal to the sample and measuring the phase shift in the response current. The impedance analysis provides information about the grain, grain boundary, electrode-sample interface, or combination of these contributions using an equivalent circuit model. The relaxation peak at low temperatures is due to immobile charge carriers, and defects or vacancies are responsible for high-temperature relaxation \cite{Nasri_CI_16}. All the samples show a negative temperature coefficient of resistance, indicating a semiconducting-type behavior, as shown in Figure~\ref{RT}. The impedance decreased with temperature due to the reduction in barrier height. At higher temperatures, the thermal energy lowers the constraints of the migration, and carriers receive the energy to flow across the grain boundary, resulting in decreased impedance. The decrease in impedance can be due to the release of immobile charges, which increase the mobility with frequency \cite{Taher_MRB_16, Idrees_JPD_10, Zhang_MRB_22}. The total complex impedance can be written as below:
\begin{equation}
Z^* = Z' +j Z''
 \end{equation}
where, the real $(Z')$ and imaginary $(Z'')$ parts of total impedance are given by
\begin{equation}
Z'=  \lvert Z \rvert cos \theta  
 \end{equation}  
\begin{equation}
Z''=  \lvert Z \rvert sin \theta  
 \end{equation}
here, $\lvert Z \rvert$ is total measured impedance, and $\theta$ is the measured phase angle (in radian). The real and imaginary impedance data show the broad relaxation peak, indicating a distribution of relaxation time where two types of relaxation peaks are observed, which are related to the grain and grain boundary contributions. The grain boundary relaxation peak occurs at the lower side of the frequency, and the grain peak occurs at the higher frequency side. These relaxation peaks shifted towards the high-frequency side, showing the reduced relaxation time with thermally activated relaxation. The asymmetric broadening shows the non-Debye type relaxation for all the samples. The non-Debye type relaxation occurs due to the presence of more than one micro-constituent having more than one relaxation time. The polycrystalline nature of the material is mainly responsible for the non-Debye-type relaxation. 

The frequency dependence of real (Z$^{'}$) and imaginary (Z$^{''}$) parts of total impedance are determined by the parallel combination of two resistance (R) and constant-phase element (CPE) related to grain and grain boundary contributions, which can be written as below \cite{Zhang_MRB_22, Karmakar_JAP_20}:  
\begin{widetext}
\begin{subequations}
\begin{equation}
\resizebox{0.94\hsize}{!}
{
$Z'=\frac{R_{G}+(R_{G}^{2}Q_{G} \omega^{\alpha}{G}) cos(\frac{\alpha{G}\pi}{2})}{[1+(R_{G}^{2}Q_{G} \omega^{\alpha}{G}) cos(\frac{\alpha{G}\pi}{2})]^{2}+[(R_{G}^{2}Q_{G} \omega^{\alpha}{G}) sin(\frac{\alpha{G}\pi}{2})]^{2}} +
\frac{R_{GB}+(R_{GB}^{2}Q_{GB} \omega^{\alpha}{GB}) cos(\frac{\alpha{GB}\pi}{2})}{[1+R_{GB}^{2}Q_{GB} \omega^{\alpha}{GB}) cos(\frac{\alpha{GB}\pi}{2})]^{2}+[R_{GB}^{2}Q_{GB} \omega^{\alpha}{GB}) sin(\frac{\alpha{GB}\pi}{2})]^{2}}$
}
\label{RZ}
\end{equation} 
 
\begin{equation}
\resizebox{0.94\hsize}{!}
{
$Z''=\frac{(R_{G}^{2}Q_{G} \omega^{\alpha}{G}) cos(\frac{\alpha{G}\pi}{2})}{[1+(R_{G}^{2}Q_{G} \omega^{\alpha}{G}) cos(\frac{\alpha{G}\pi}{2})]^{2}+[(R_{G}^{2}Q_{G} \omega^{\alpha}{G}) sin(\frac{\alpha{G}\pi}{2})]^{2}}+
\frac{(R_{GB}^{2}Q_{GB} \omega^{\alpha}{GB}) cos(\frac{\alpha{GB}\pi}{2})}{[1+R_{GB}^{2}Q_{GB} \omega^{\alpha}{GB}) cos(\frac{\alpha{GB}\pi}{2})]^{2}+[R_{GB}^{2}Q_{GB} \omega^{\alpha}{GB}) sin(\frac{\alpha{GB}\pi}{2})]^{2}}$
} 
\label{IZ}
 \end{equation}
\end{subequations}
\end{widetext}
Here, Q is an independent parameter called capacitance of CPE as determined by C= R$^{1-n}$Q$^{\frac{1}{n}}$, and $n$ is the exponential factor which determines the non-ideal type behavior of the material. The $n$ has values between 0 to 1; where $n=$ 1 for the ideal capacitor and $n=$ 0 for the ideal resistor \cite{Karmakar_JAP_20}. The fitting of real Z$^{'}$ and imaginary Z$^{''}$ parts of impedance data using equations \ref{RZ} and \ref{IZ} are shown in Figure \ref{ReZ} and Figure \ref{ImZ}. We observe that the grain boundary is more resistive and capacitive than grains, giving the larger values of relaxation times appear at the lower side of the frequency spectrum. The larger values of grain boundary resistance are due to the trapping of charge carriers at the boundary interface due to defects, oxygen vacancies, dangling bonds, and lattice strain. The value of grain and grain boundary resistance decreases with temperature due to a reduction in potential barrier and an increase in activation energy. The increase in grain and grain boundary capacitance is due to ionic and electronic charge density and the release in interfacial polarization. The analysis (determined using the values of $\alpha$) shows the grain and grain boundary relaxation of non-Debye nature in the measured frequency and temperature range.

\begin{figure*}
    \centering
    \includegraphics[width=0.80\textwidth]{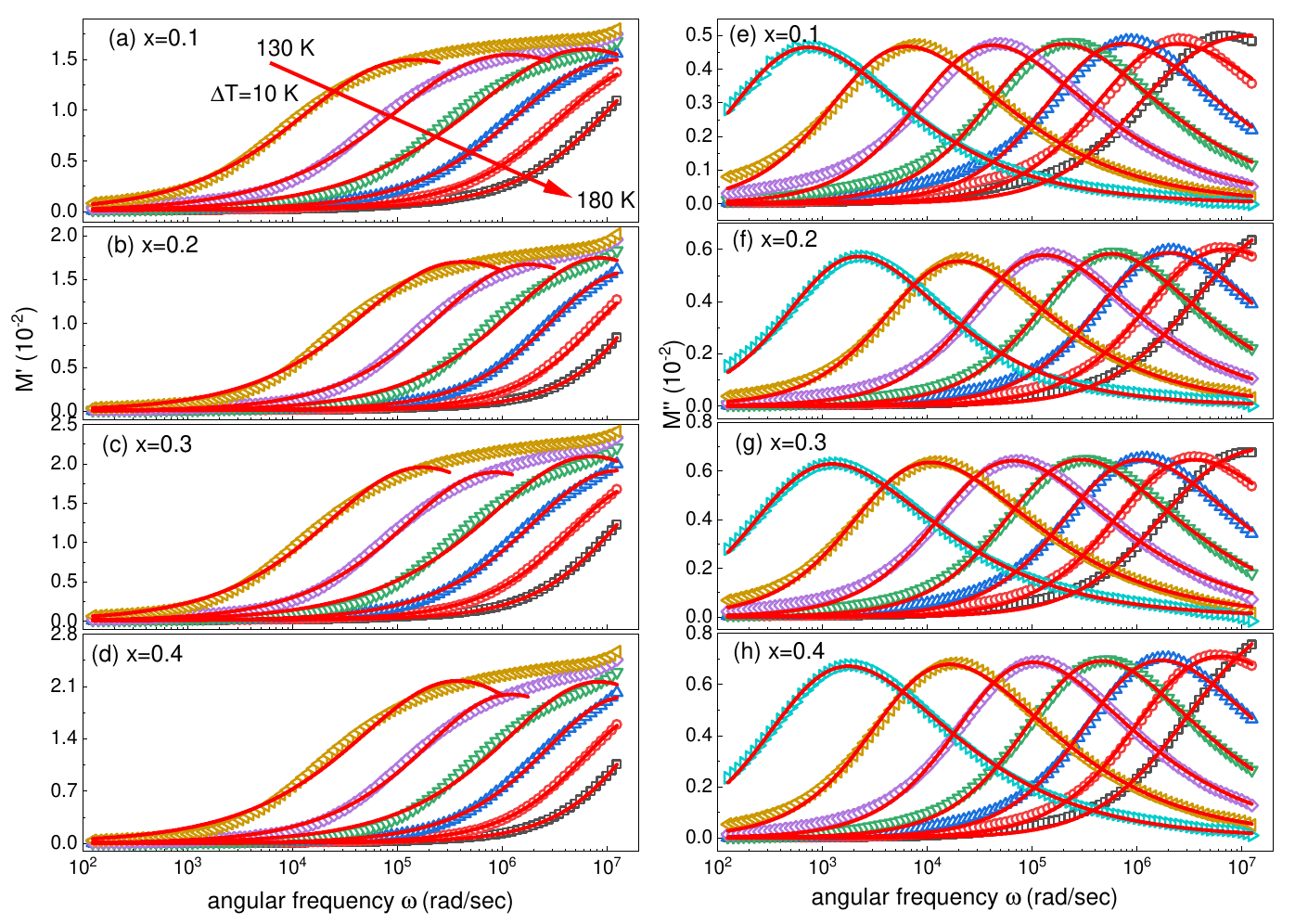}
    \caption{The temperature dependent variations in real (M$^{'}$) and imaginary modulus (M$^{''}$) at selected temperatures for the Na$_{3}$Zr$_{2-x}$Ti$_{x}$Si$_2$PO$_{\rm 12}$ ($x=$ 0.1--0.4) samples. Here, the open symbol represents the measured experimental data, and the solid line is fit using equation \ref{M}.}
\label{M'M''}
 \end{figure*}

Next, we will study the electric modulus as a function of temperature and frequency. Macedo introduced the concept of electric modulus study to understand the relaxation dynamics, conductivity mechanism, ion hopping rate, and distribution of relaxation times under the influence of an external field while electric displacement remains constant \cite{Macedo_PCG_72}. The electric modulus studies have an advantage as they eliminate the electrode polarization effect and differentiate the space charge effect from the bulk. It exhibits the largest peak in the imaginary modulus (M$^{''}$) as a function of frequency. The modulus study is applicable for conducting, nonconducting, and ionic conducting samples \cite{Nasri_CI_16, Baskaran_JAP_02, Macedo_PCG_72}. The complex electric modulus (M$^{*}$) is the reciprocal of complex permittivity representing the real dielectric relaxation dynamics is represented as follows: 
\begin{equation}
M^{*}(\omega)=\frac{1}{\epsilon^{*}(\omega)}=M'(\omega)-jM''(\omega)= j\omega C_{0} Z^{*}
 \end{equation}
The real and imaginary parts of the electric modulus can be written as follwoing: 
\begin{subequations}
\begin{equation}
M^{'}(\omega)= \omega C_{0} Z^{''}
\label{RM}
 \end{equation}
\begin{equation}
M^{''}(\omega)= \omega C_{0} Z^{'}
\label{IM}
 \end{equation}
\end{subequations}
Here, $\omega$ is the angular frequency, C$_{0}$ is the vacuum capacitance as given by C$_{0}$=$\epsilon$$_{0}$ $\frac{A}{d}$, Z$^{'}$ and Z$^{''}$ are the real and imaginary parts of impedance. The real and imaginary parts of the electric modulus as calculated using equations \ref{RM} and \ref{IM} are shown in Figures~\ref{M'M''}(a--d) and Figures~\ref{M'M''}(e--h), respectively. It shows a negligible magnitude at lower frequency values and an abrupt increase in magnitude for higher frequency confirming the capacitive nature of all the samples. The smaller magnitude of (M$^{'}$) for lower frequencies is due to lack of restoring force that governs the required mobility of relaxation, excluding the electrode polarization contribution. The abrupt increase shifts towards the higher frequency side with an increase in temperature, showing the temperature-dependent relaxation behavior \cite{Lakhdar_MSSP_15}. The dispersive nature of M$^{'}$ is due to conductivity relaxation and short-range mobility. The relaxation peak in imaginary modulus spectra M$^{''}$ shifts towards the higher frequency side with an increase in temperature, suggesting the ionic nature of these samples \cite{Taher_MRB_16, Taher_Ionics_15}. The height of the relaxation peak is nearly the same suggesting the weakly temperature-dependent capacitance. The relaxation peak in modulus spectra occurs due to the accumulation of charge at the sample-electrode interface, which is called as interfacial polarization. The relaxation peak at lower temperatures is due to contributions from grains, and the high-temperature relaxation peak is due to grain boundary contributions. The accumulation of charges increases with an increase in temperature, shifting the relaxation peak towards the higher frequency side. The frequencies below the modulus peak show the long-range hopping of charge carriers, while above-peak frequency carriers follow the short-range hopping \cite{Nasri_CI_16, Singh_JALCOM_17, Sondarva_JALCOM_21}.  

The dielectric relaxation is understood using Laplace transformation of Kohlraush-Williams-Watts (KWW) decay function in the time domain given by $\varphi$(t) = exp (-$\frac{t}{\tau}$)$^{\beta}$, where $\varphi$(t) represents the decay of electric field inside the material and $\beta$ is an exponent varies between 0 to 1 decides the relaxation-time distribution \cite{Moynihan_PCG_73}. Here, $\beta$=1 represents the ideal Debye type relaxation having no interactions among charge carriers, and $\beta$=0 shows the maximum interaction among dipoles. The $\tau$ represents the relaxation time of charge carriers. The KWW function relates  the decay of the electric field within the dielectric material. The KWW function was modified to determine the modulus relaxation in the frequency domain, where the M$^{''}$ can be written as \cite{Taher_MRB_16, Bergman_JAP_00} 
\begin{equation}
\label{eq-M}
    M''=\frac{M''_{max}}{[(1-\beta)+(\frac{\beta}{1+\beta})[\beta(\frac{\omega_m}{\omega})+(\frac{\omega}{\omega_m})^\beta]}
\end{equation}
Here, M$^{''}$$_{max}$ is the peak maximum of imaginary modulus, and $\omega$$_{m}$ is the angular frequency for peak modulus. The peak frequency in the modulus relaxation increases towards the high-frequency side further confirming the ionic nature of these samples. The relaxation peaks, as shown in Figure \ref{M'M''}(e--h), could not be fitted using the  Bergman relaxation model due to their asymmetric nature and shows the non-Debye type relaxation for all the samples. The real ($M'$) and imaginary ($M''$) parts of electric modulus are fitted using the Havriliak-Negami (HN) relaxation model to understand the non-Debye type relaxation dynamics,  given as below \cite{Nasri_CI_16, Pal_JAP_19}: 
\begin{equation}
M^*(\omega)=M_{\infty} + \frac{(M_{s}-M_{\infty})}{[1+(i\omega \tau)^{\alpha}]^{\gamma}}
\label{M}
 \end{equation}
where, the M$_{s}$ and M$_{\infty}$ are the static and high-frequency limit of complex electric modulus, $\alpha$ and $\gamma$ are the shape parameters characterize the nature of the electric modulus having the values between 0 to 1, and $\tau$ is the relaxation time. The real and imaginary components of Havriliak-Negami (HN) equation \ref{M}, as below \cite{Islam_MRB_23, Alvarez_PRB_91, Pal_JAP_19}
\begin{subequations}
\begin{equation}
M^{'}=M_{\infty} + \frac{(M_{s}-M_{\infty}) ~ Cos(\gamma \phi)}{[1+2(\omega\tau)^{\alpha} ~ Cos (\frac{\pi \alpha}{2})+(\omega\tau)^{2\alpha}]^{\frac{\gamma}{2}}}
\label{M'}
 \end{equation}  
\begin{equation}
M^{''}= \frac{(M_{s}-M_{\infty}) ~ Sin(\gamma \phi)}{[1+2(\omega\tau)^{\alpha} ~ Cos (\frac{\pi \alpha}{2})+(\omega\tau)^{2\alpha}]^{\frac{\gamma}{2}}}
\label{M''}
 \end{equation}  
\end{subequations}
 where 
\begin{equation}
\phi= \arctan [\frac{(\omega\tau)^{\alpha} ~ Sin (\frac{\pi \alpha}{2})}{1+(\omega\tau)^{\alpha} ~ Cos (\frac{\pi \alpha}{2})}]
\end{equation}
The real and imaginary parts are fitted using equations \ref{M'} and \ref{M''}, respectively, as shown in Figure~\ref{M'M''} by considering the M$_{s}$, M$_{\infty}$, $\tau$, $\alpha$ and $\gamma$ as unknown parameters. The values of $\alpha$ and $\gamma$ are in the range of 0.34--0.79 and 0.28--0.76, respectively, further confirming the non-Debye type relaxation at lower temperatures. \cite{ Alvarez_PRB_91, Pal_JAP_19}.

The relaxation frequency increases with temperature due to increased thermal activation of charge carriers, resulting in a decreased relaxation time \cite{Taher_MRB_16, Singh_JALCOM_17}. The shifting of relaxation frequency towards higher value suggests the modulus relaxation as a thermally activated process. The activation energy of modulus relaxation (E$_{M}$) is determined by the Arrhenius law of thermal activation: 
\begin{equation}
    \label{omega-Arr}
    \omega_{m} = \omega_0 ~ exp (\frac{-E_M}{k_BT})
\end{equation}
Here, $\omega_0$ is the pre-exponential factor of frequency, k$_B$ is the Boltzmann constant, T is the measured temperature, and $E_M$ is the activation energy of the modulus relaxation. The modulus activation energy is determined using the linear least square fit of ln$(\omega_{m})$ versus (1000/T), as shown in Figure~3 of \cite{SI_RCM_24}, which shows the nearly similar activation energy around 0.27~eV and indicates a similar relaxation process for all the samples. The magnitude of activation energy suggests the conduction is due to the hopping mechanism. To determine the modulus relaxation type and its temperature dependence, modulus scaling is performed, as shown in Figure 4 of \cite{SI_RCM_24}. Here, the y-axis is scaled using the peak modulus (M$^{''}$$_{max}$), and the x-axis is scaled using the corresponding peak angular frequency ($\omega$$_{max}$) for all the samples. The merging of all curves over each other indicates a similar type of relaxation independent of temperature. The merging of all curves confirms the relaxations occurring at different time scales have similar activation energy \cite{Kaswan_JALCOM_21, Hadded_RSC_20}.

\begin{figure}
    \centering
    \includegraphics[width=0.47\textwidth]{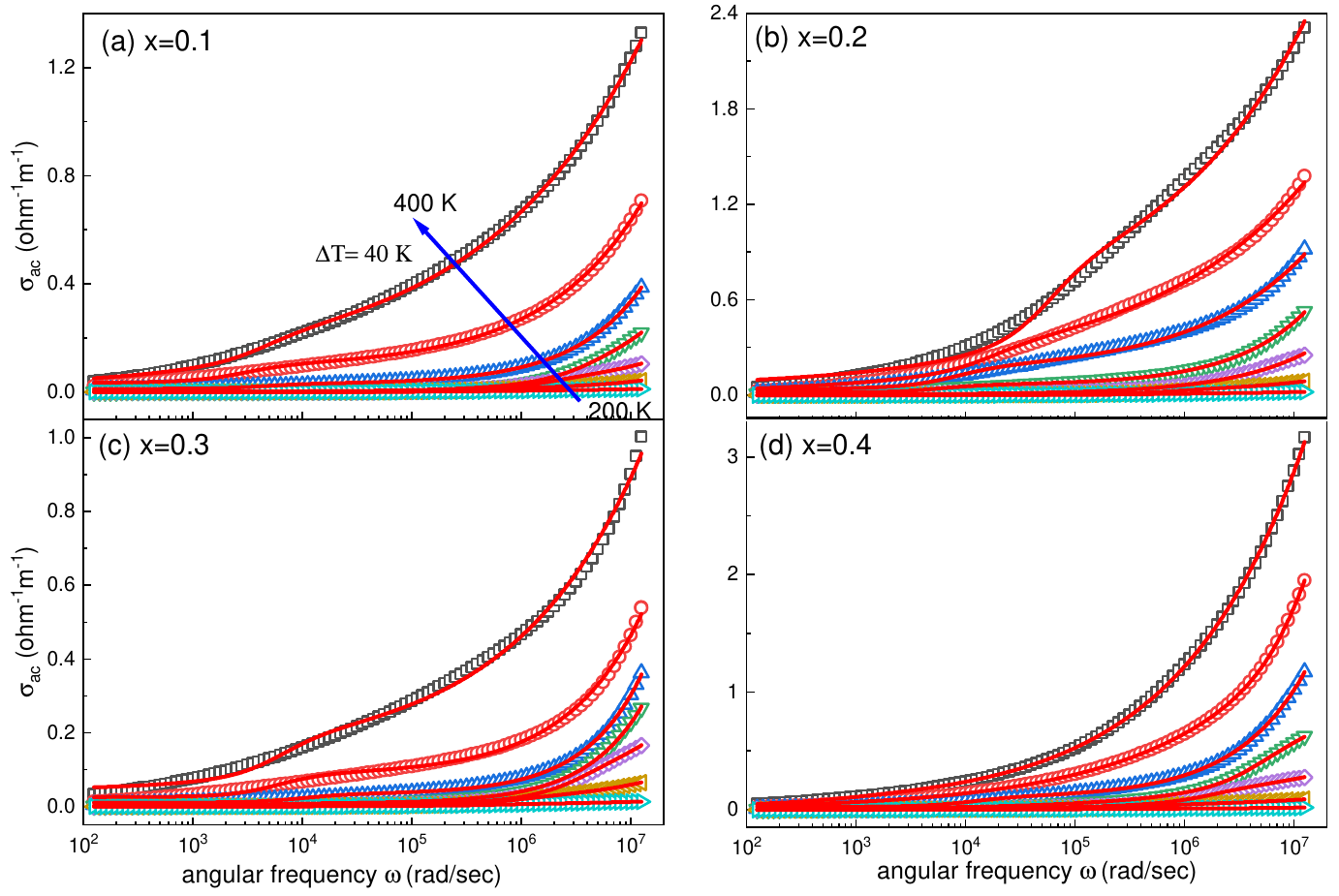}
    \caption{The {\it a.c.} conductivity variations of Na$_{3}$Zr$_{2-x}$Ti$_{x}$Si$_2$PO$_{\rm 12}$ ($x=$ 0.1--0.4) samples in the temperature range of 200--400 K. The experimental data (open symbol) are fitted using the modified power using equation~\ref{MPLT}.}
\label{ST}
 \end{figure}

The electrical conduction in an alternating field is the contribution from the charge carriers in phase with the applied external electric field. We have measured the {\it a.c.} conductivity in the frequency range of 20 Hz -- 2 MHz and in the temperature range of 200 K-440 K. The a.c. conductivity is the contribution from the carriers in close vicinity of the Fermi level; it is a combined effect of carrier relaxation, hopping, and diffusion and the total conductivity is the contribution of free carriers or bound charge carriers, which can be determined using the frequency dependence; if the conductivity increases (decreases) with an increase in frequency, then the contribution comes from bound charge carriers (free charge carriers) \cite{Sumi_JAP_10}. In an external alternating field, the charge carriers move from one state to another via a hopping mechanism. At lower frequency values, the charge carriers hop for longer distances due to considerably larger time period available for carriers, while short-range hopping occurs at higher frequencies. The conductivity increases abruptly above a characteristic frequency called hopping frequency due to short-range hopping dominates, and in this case, the capacitance impedance becomes lower than the resistor impedance. The conductivity increases with temperature due to higher thermal activation of charge carriers resulting in the larger hopping. Figure~\ref{ST} shows the {\it a.c.} conductivity calculated using the equation below \cite{Raut_JAP_18, Nasri_CI_16, Karmakar_JPCM_19} 
\begin{equation}
\sigma_{ac}(\omega)= \frac{d}{A} \left[ \frac{Z'}{Z'^2+Z''^2} \right]
\end{equation}
Here, $d$ is the thickness of the sample, A is the electrode area, Z$^{'}$ and Z$^{''}$ are the real and imaginary parts of the total impedance. We find that the conductivity increases with frequency, confirming the contributions from bound charge carriers. The {\it a.c.} conductivity analysis shows two types of behavior: (a) a broad plateau or nearly constant region at lower frequencies and (b) a highly dispersed region at higher frequencies. The frequency-dependent relaxation behavior is explained using Funke's model \cite{Funke_PSSC_93, Sumi_JAP_10}; according to this model, the charge carriers migrate for a larger range at lower frequencies due to longer time period available for the carriers, which gives the frequency-independent conductivity. Moreover, two competing processes occur at higher frequencies: (a) successful hopping and (b) unsuccessful hopping due to the shorter periods available for the carriers. The ratio of these two competing processes gives the dispersed behavior in {\it a.c.} conductivity. The changes from frequency-independent to frequency-dependent regions vary from long-range hopping to short-range hopping \cite{Sharma_JMS_20, Nasri_CI_16}. The conventional frequency dependence of {\it a.c.} conductivity  is analyzed using the universal power law called as Jonscher's law, given as below \cite{Jonscher_Nature_77, Nasri_CI_16, Raut_JAP_18}   
\begin{equation} 
\label{acsig}
	\sigma_{ac} (\omega) = \sigma_{dc} +A ~\omega^s
	\end{equation}
Here, $\sigma$$_{ac}$ is the measured ac conductivity, $\sigma$$_{dc}$ is the frequency independent part called {\it d.c.} conductivity, $A$ is the parameter that determines the strength of polarization, and $s$ is the parameter that determines the interaction between the ion and surrounding medium. 

Interestingly, the $s$ parameter is temperature and frequency-dependent, which determines the conduction mechanism in the measured temperature range.  If $s$ is a constant of value around 0.8, or increased slightly with temperature, the conduction mechanism is governed by the quantum mechanical tunneling (QMT) \cite{Ghosh_PRB_90}. If $s$ increases (decreases) linearly with temperature, the conduction is governed by non-overlapping small polaron hopping (NSPT) [correlated barrier hopping (CBH)] \cite{Mollh_JAP_93, Ghosh1_PRB_90}. If $s$ is a value equal to unity at room temperature and decreases with the temperature reaches a minimum and further increases with temperature, the conduction mechanism is governed by the overlapping large-polaron hopping \cite{Long_AIP_82, Kahouli_JPCA_12}. We could not fit the {\it a.c.} conductivity data using the universal power law, as mentioned in the equation \ref{acsig}; the conductivity data are fitted using the modified power law \cite{Bechir_JAP_14, Karmakar_JPCM_19, Megdiche_JALCOM_14}
\begin{equation} 
\label{MPLT}
\sigma_{(ac)} (\omega) = \frac{\sigma_l}{(1+\omega^2 \tau^2)}+\frac{\sigma_{h} \tau^2 \omega^2 }{1+\omega^2 \tau^2} +A ~\omega^s
\end{equation}
where, $\sigma_l$ and $\sigma_{h}$ represent the conductivity measurement at the lowest and highest frequency, $\tau$ is the characteristics relaxation time, $A$ and $s$ have the same meaning as in universal power law. The fitted results using the equation \ref{MPLT} are shown in Figure~\ref{ST}. The {\it a.c.} conductivity data are well-fitted in the measured frequency and temperature range using the modified power law.  The temperature dependence of the $s$ parameter is shown in Figure 5 of \cite{SI_RCM_24} for all the samples. It is observed that the $s$ parameter initially increases with an increase in temperature, suggesting the NSPT type conduction at lower temperatures. In this type of conduction charge carriers tunnels near the Fermi level. The small polaron means the distorted localized regions that do not overlap \cite{Kahouli_JPCA_12, Mathlouthi_JALCOM_19}. The analysis shows the change in conduction mechanism around 320 K, where $s$ parameter start decreasing with temperature. This type of variations suggests the CBH type conduction above 320 K. In this type of conduction mechanism, the charge carriers hop between two defect states separated by a distance of R$_{\omega}$ with a separation of Coulomb potential energy W$_{M}$. In the CBH model, hopping occurs due to single or bipolaron hopping by creating the disorder in the surrounding medium, producing structural-type defects \cite{Taher_MRB_16, Mollh_JAP_93}.

\section{\noindent ~Conclusions}

The Na$_{3}$Zr$_{2-x}$Ti$_{x}$Si$_2$PO$_{\rm 12}$ ($x=$ 0.1--0.4) bulk samples were prepared using the solid-state reaction method. The XRD analysis using the Rietveld refinement method shows the monoclinic phase (space group C 2/c). The resistivity measurement shows the semiconducting type behavior above the room temperature with a change in slope around 400 K, indicating the possibilty of monoclinic to rhombohedral structural phase transition. The electric permittivity ($\epsilon_{r}$)  shows the increasing behavior having the highest permittivity of the order of 10$^{7}$ at 440 K.  The electric permittivity's frequency and temperature dependence are explained based on space charge polarization or interfacial polarization and Maxwell-Wagner relaxation mechanisms. The dielectric loss behavior shows the double relaxation peaks having the Arrhenius type thermally activated relaxation. The magnitude of relaxation activation energy shows similar types of relaxation for all the samples. The real ($\epsilon_{r}$$^{'}$) and imaginary ($\epsilon_{r}$$^{''}$) parts of permittivity data were fitted using the modified Cole-Cole equation, including the conductivity term, which shows the non-Debye type relaxation in the measured frequency and temperature range. The real (Z$^{'}$) and imaginary (Z$^{''}$) impedance analysis shows the relaxation due to grain and grain boundary contributions of non-Debye nature. The electric modulus data were analyzed using the Hervialik-Negami (HN) equation, which confirmed the non-Debye type relaxation. The imaginary modulus scaling analysis shows a similar type of relaxation independent of temperature. The {\it a.c.} conductivity behavior, fitted using the modified power law, is found to increase with temperature, and frequency. The temperature dependence of the $s$ parameter shows the change in the conduction mechanism from non-overlapping small polaron (NSPT) to correlated barrier hopping (CBH) in the studied temperature range. \\

\section{\noindent ~Acknowledgments}

RM thanks IUAC for providing the experimental facilities and Department of Physics of IIT Delhi for providing XRD facility. RSD acknowledges the financial support from SERB-DST through a core research grant (project reference no. CRG/2020/003436).

\end{document}